# Psychological and behavioural responses in human-agent vs. human-human interactions: a systematic review and meta-analysis

Submitted for peer-review

Jianan Zhou[1*], Fleur Corbett[1], Joori Byun[2], Talya Porat[1], Nejra van Zalk[1]
[1]Dyson School of Design Engineering, Imperial College London, London, United Kingdom
[2]Independent Researcher

[*]**Correspondence:** Correspondence concerning this article should be addressed to Jianan Zhou, Dyson School of Design Engineering, Imperial College London, London, United Kingdom. Email: jianan.zhou22@imperial.ac.uk

**Abstract**
Interactive intelligent agents are being integrated across society. Despite achieving human-like capabilities, humans' responses to these agents remain poorly understood, with research fragmented across disciplines. We conducted a first systematic synthesis comparing a range of psychological and behavioural responses in matched human-agent vs. human-human dyadic interactions. A total of 162 eligible studies (146 contributed to the meta-analysis; 468 effect sizes) were included in the systematic review and meta-analysis, which integrated frequentist and Bayesian approaches. Our results indicate that individuals exhibited less prosocial behaviour and moral engagement when interacting with agents vs. humans. They attributed less agency and responsibility to agents, perceiving them as less competent, likeable, and socially present. In contrast, individuals' social alignment (i.e., alignment or adaptation of internal states and behaviours with partners), trust in partners, personal agency, task performance, and interaction experiences were generally comparable when interacting with agents vs. humans. We observed high effect-size heterogeneity for many subjective responses (i.e., social perceptions of partners, subjective trust, and interaction experiences), suggesting context-dependency of partner effects. By examining the characteristics of studies, participants, partners, interaction scenarios, and response measures, we also identified several moderators shaping partner effects. Overall, functional behaviours and interactive experiences with agents can resemble those with humans, whereas fundamental social attributions and moral/prosocial concerns lag in human-agent interactions. Agents are thus afforded instrumental value on par with humans but lack comparable intrinsic value, providing practical implications for agent design and regulation.

**Introduction**
Interactive intelligent agents, such as chatbots, virtual humans, and robots, are increasingly embedded across professional and personal domains, including healthcare, education, business, and leisure[1]. Recent breakthroughs in artificial intelligence (AI) have advanced these systems towards, and in some cases beyond, human-level performance on specific tasks[2–5]. Leading technology labs like Google's DeepMind suggest no inherent technical limit to these systems achieving human-level capabilities across various domains[6]. Evolving from narrow applications to broader competence, these systems move beyond reactive "tools" or "assistants" to proactive "agents": autonomous computational systems capable of goal-directed behaviour, interaction with the environment, and task execution with minimal human involvement[7–10]. Human interaction with these agents fundamentally differs from traditional human-computer interaction (HCI)[11]. Agents are assuming roles primarily reserved for humans, serving as companions, collaborators, advisors, and even opponents[12–15]. Scenarios such as cooperation, competition, and strategic encounter, originally confined to human-human interaction (HHI), now emerge in human-agent interaction (HAI)[16–19]. These developments are blurring long-standing boundaries between humans and machines.

As people engage with intelligent agents in socially complex settings requiring psychological attunement[20–22], a key question emerges: can agents' human-like performance evoke the same psychological and/or behavioural responses—the way individuals feel, think and act within an interaction[23]—as those with human partners? This knowledge gap calls for rigorous investigation. For example, individuals may or may not maintain comparable *task performance* when collaborating with agents as they do with humans. Their sense of *personal agency* may vary during agent interaction. Core *social responses*, such as alignment, trust, and morality, may be inherently tied to human presence or may be elicited by agents displaying human-like behaviour. Insights into the transferability of human psychology and behaviour from HHI to HAI is critical to inform responsible agent design and regulation[24–27]. It is thus important to compare a range of psychological and behavioural responses in HAI vs. HHI, where the task performance of agent and human partners is matched, and to further explore the conditions under which similarities and differences emerge.

Empirical evidence on human responses in HAI vs. HHI is fractured across disciplinary silos and methodological traditions. The rise of interactive intelligent agents has attracted scholars beyond computer science and robotics, including psychology, communication, business, and marketing, spurring a proliferation of related studies[28]. Nevertheless, inconsistent constructs, divergent methodologies, and isolated investigations have impeded a cohesive understanding of response comparability, yielding largely mixed findings. Whereas some studies report equivalent responses to agent and human partners, others reveal substantial deficits or qualitatively distinct responses. This inconsistency exists under ostensibly similar contexts, without consensus on whether differences stem from research designs, agent types, interaction tasks, or participant demographics.

The mixed empirical evidence is mirrored by a diversity of theoretical perspectives. For example, the Computers Are Social Actors paradigm and its predecessor, the Media Equation, propose that individuals respond to intelligent agents with minimal human-like cues as they would to humans, suggesting that social responses in HAI mirror those in HHI[29,30]. This proposition of equivalence, however, is challenged by other frameworks. The Threshold Model of Social Influence draws a key distinction: when behaviour appears equally realistic, individuals show stronger deliberate social responses when they believe they are with humans rather than agents in virtual settings—though automatic, low-level reactions remain unchanged across both[31]. Theories of anthropomorphism shift focus to the attribution of human-like characteristics to agents' real or imagined behaviour[32,33]. It has been suggested that anthropomorphising behaviour (i.e., the observable ways in which individuals respond to agents as they would to humans) should be studied in HAI without presuming response equivalence to HHI[34]. Theories focusing on social perceptions of agents provide additional insights. For example, the Modality, Agency, Interactivity, and Navigability model proposes that perceived agency (human vs. algorithm) activates different cognitive heuristics influencing perceived credibility[35]. The Social Presence Theory examines the degree to which agents are perceived as real social actors[36], with a recent meta-analysis showing how social cues in agents increase perceived social presence[37]. The Uncanny Valley Theory introduces a critical caveat regarding human-like design, warning that near-human realism with subtle imperfections evokes eeriness and discomfort[38]. This abundance of existing theoretical frameworks underscores the need for a cross-disciplinary, systematic understanding of human responses in HAI vs. HHI, and identifying factors influencing their similarities or differences.

Four previous reviews have addressed related topics. One narrative review[25] argued that humans engage with agents in ways resembling HHI, including relationship building, suggesting that similarities outweigh differences and that theories of HHI apply to HAI. By contrast, another narrative review[27] cautioned against equating HAI with HHI, arguing that current agents lack key social affordances underpinning HHI theories and urging the development of HAI-specific models. A third integrative review[39] of trust in HAI vs. HHI found similar developmental processes but differences in expression and calibration, and suggested narrowing gaps between perceptions of humans and agents to improve user trust. Nevertheless, there are two primary limitations with these reviews. They prioritised argument-driven theoretical integration over evidence-driven systematic synthesis. Also, they lack focus on literature directly comparing responses between HAI and HHI, which, even when revealing response similarities, only metaphorically support the Media Equation[40]. A fourth review and meta-analysis[41] systematically compared responses in HAI and HHI by specific partner types (virtual agents vs. avatars) and found that avatars exerted stronger social influence than agents. While insightful, this review collapsed responses (e.g., subjective perceptions, affects, task performance, physiological measures) into uniform so-called measures of social influence and pooled them in one meta-analysis. Although most psychological and behavioural responses in HAI/HHI fall under the umbrella of social responses[42], treating them as a monolithic entity obscures conceptual distinctions. Thus, no reviews to date provide a holistic picture of how various types of human responses compare between HAI and HHI.

To address prior limitations, we conducted a systematic review and meta-analysis of individuals' psychological and behavioural responses in dyadic interactions with functionally equivalent (i.e., performance-matched) agent vs. human partners, quantifying the effects of partner type (agent vs. human; hereafter, *partner effects*) on specific responses and exploring moderators of these partner effects. We aimed to answer four research questions:
- RQ1: Which psychological and/or behavioural responses have been investigated in HAI compared to HHI?
- RQ2: Which specific response types differ between HAI and HHI?
- RQ3: Which specific response types are similar between HAI and HHI?
- RQ4: To what extent do study, participant, partner, scenario, and measurement characteristics moderate partner effects on different response types?

Aligning with a human-centred perspective, we treated "agent" as a functional metaphor rather than a strict technical term[7]. A computational system was considered an intelligent *agent* if it assumed a human-equivalent role while matching the task performance of its human counterpart, irrespective of technical implementation. Agent architectures may include traditional machine learning, rule-based algorithms, Wizard-of-Oz setups, or more recent generative AI (GenAI). In addition, the term "interaction" is central to HAI/HCI, yet remains overloaded and ambiguous[43]. It has been understood in various ways: as an experiential stream of subjective expectations, experiences, and memories[44]; as a system's disposition for interaction from a design perspective[45,46]; or as a process involving mutual exchanges[47]. In this review, we define *interaction* as the engagement between participants and partners for some purpose within certain contexts, involving reciprocal information exchange or unidirectional transmission with at least one party taking an active role. This requires participatory engagement, where participants act as interactants rather than observers in real-time or hypothetical (i.e., vignette-based imaginative) scenarios. Given critiques of understanding interaction as mutual exchanges in HCI[43], we also included unidirectional interactions—for example, participants speak while partners listen or vice versa—provided partners are not presented as passive prerecorded stimuli but possessing interactability (i.e., the disposition and readiness to engage)[48]. Furthermore, mixed empirical evidence on human responses in HAI vs. HHI implies the existence of moderating effects. We chose potential moderators based on a prior meta-analysis[41] on the effects of agents vs. avatars and on a conceptual framework[49] of social responses to agents, which highlights the roles of agent features, individual differences, and the interaction context in shaping whether individuals respond similarly to agents and other humans. We thus examined factors across characteristics of partners, participants, and interaction scenarios. We also explored factors related to study design and response measures, as methodological variability could influence partner effects.

**Results**

The study selection process is summarised in the PRISMA flowchart (Fig. 1). After title-abstract screening (11,456 records) and full-text screening (815 articles), 162 studies (from 122 articles) met our eligibility criteria (Supplementary Table 1). Key characteristics of eligible studies are provided in Supplementary Table 2. These studies investigated a broad set of psychological and behavioural responses, which we classified into distinct response types and meta-analysed separately. Sixteen studies examined rare responses with insufficient data for meta-analysis[50], and these were narratively synthesised in Supplementary Table 3. The final quantitative synthesis included 146 studies (468 effect sizes; 3.21 per study).

Integrating frequentist and Bayesian approaches, we conducted a series of meta-analyses comparing individuals' psychological and behavioural responses in dyadic interactions with performance-matched agent vs. human partners. Studies included in the meta-analyses were published between 2003 and 2024. The average sample size was 163.73 ($SD$ = 163.19; range = 8–945). For seven studies that reported only total sample sizes and included conditions beyond HAI/HHI, we assumed equal condition sizes[51,52]; removing these yielded a similar average of 164.53 ($SD$ = 166.27). Mean participant age was 28.80 years ($SD$ = 8.64, range = 18.60–78.79), and samples averaged 55.96% female ($SD$ = 17.45%, range = 0–100%).

**RQ1: Which psychological and/or behavioural responses have been investigated in HAI compared to HHI?**

We identified 23 types of human responses, each with at least five studies for meta-analysis[50]. They were grouped into six themes: prosociality and morality, social perceptions of interaction partners, trust in interaction partners, social alignment with interaction partners, personal agency and task performance, and interaction experiences. An overview of the response types, grouped under these themes and with detailed conceptualisations, is provided in Table 1.

To derive response types and themes, we performed a posteriori classification of the diverse human responses investigated across studies. After a conceptual-to-empirical approach[53] to classification development failed to align with extracted responses (see Supplementary Note 1), we adopted an empirical-to-conceptual approach[53], allowing the classification scheme to arise directly from the dataset. We reviewed the descriptions and measures of all human responses extracted from the studies and inductively classified them into distinct response types. Conceptually aligned response types were further grouped into six identified themes, with a residual category for unclassified responses.

**RQ2: Which specific response types differ between HAI and HHI?**
Meta-analyses revealed notable differences in prosociality and morality, as well as in social perceptions of partners, between HAI and HHI. Detailed results are summarised in Table 2, with pooled effect sizes visualised in Fig. 2.

**Prosociality and morality.** *Prosocial behaviour* refers to voluntary actions intended to benefit others[54], with frequentist meta-analysis revealing a medium-to-large partner effect. Individuals behaved significantly less prosocially when interacting with agent vs. human partners, for example by sharing less in dictator games or adapting less to the partner's perspective in joint tasks. *Moral engagement,* the psychological and behavioural commitment to moral standards[55,56], showed a small-to-medium partner effect. Individuals engaged significantly less morally with agent vs. human partners, with reduced moral acts, intentions, or feelings of guilt.

**Social perceptions of interaction partners.** On average, *perceived social presence*—the sense of the partner being "there" and socially real—showed a small partner effect, with agents perceived as significantly less socially present than human partners. For two core dimensions of social perception[57], *perceived likeability* reflects affective evaluations of the partner[58,59] (e.g., warmth, friendliness), while *perceived competence* captures evaluations of capability[58,60] (e.g., intelligence, effectiveness). Both showed small-to-medium partner effects, with agents perceived as significantly less likeable and competent than human partners. The agent disadvantage was greater for higher-order social constructs: a medium-to-large partner effect on *agency attribution*—capacity for intentional action[61]—and a medium effect on *responsibility attribution*—accountability for outcomes[62], with agents attributed significantly less agency and responsibility.

Bayesian meta-analytic results were consistent with frequentist estimates. Bayesian support was decisive for partner effects on prosocial behaviour and moral engagement. Regarding social perceptions, support was substantial for partner effects on perceived social presence, strong for agency attribution, very strong for perceived likeability, and decisive for perceived competence and responsibility attribution.

**RQ3: Which specific response types are similar between HAI and HHI?**
Meta-analyses revealed general response similarities between HAI and HHI across four response themes: trust, social alignment, perceived agency and task performance, and interaction experiences. Detailed results are provided in Table 2 and Fig. 2.

**Trust in interaction partners.** Trust is "a willingness to be vulnerable" in the absence of the ability to monitor the trustee[63], encompassing both behavioural and psychological aspects. Frequentist meta-analyses revealed no significant differences in *behavioural trust* (e.g., risk-taking or reliance on the partner under uncertainty) and *subjective trust* (e.g., perceived trustworthiness or reliability) towards agent vs. human partners.

**Social alignment with interaction partners.** *Social alignment* refers to the voluntary alignment of internal states and behaviours with the partner[64], such as self-other integration and behavioural synchrony. Like behavioural trust, it is critical for interaction coordination[64] and showed no significant difference with agent vs. human partners.

**Personal agency and task performance.** *Perceived self-agency* and *self-disclosure* are self-oriented processes[65] during interaction. Self-agency is the sense of control over one's own body and external events, while self-disclosure involves sharing personal information. Both showed no significant differences during agent vs. human interactions. Moreover, *strategic economic behaviour*—deliberate choices in economic games while acknowledging the interdependence of players' actions[66]—did not differ significantly by partner type. *Objective task performance*—quantifiable assessments of task execution (e.g., accuracy or response time)—also did not differ significantly.

**Interaction experiences.** On average, *perceived relational qualities* of agent vs. human partners—subjective evaluations of relational attributes the partner exhibits during interaction[67], such as rapport and empathy—did not differ significantly. Likewise, no significant differences were found for *affective valence* (e.g., unpleasantness–pleasantness) and *arousal* (e.g., calmness–excitement). *Interaction satisfaction*—overall evaluation of how well the interaction meets one's needs and expectations[68]—did not differ significantly by partner type, nor did *future interaction intention* (e.g., willingness to engage again), *perceived naturalness*, or *enjoyment of interaction*. For subjective task experience, *subjective workload* (e.g., perceived effort) and *task engagement* (e.g., enjoyment or immersion in the task) did not differ significantly.

Bayesian meta-analytic results were consistent with frequentist estimates. However, Bayesian evidence was ambiguous for the absence of partner effects on subjective trust, affective valence, perceived interaction naturalness and enjoyment. For other response types, evidence for the absence of partner effects was clearer. Bayesian support was strong for behavioural trust, social alignment, perceived self-agency, self-disclosure, and objective task performance, and substantial for strategic economic behaviour and most interaction experiences, including perceived partner relational qualities, interaction satisfaction, future interaction intention, affective arousal, subjective workload, and task engagement.

**RQ4: To what extent do study, participant, partner, scenario, and measurement characteristics moderate partner effects on different response types?**
We conducted univariate meta-regressions exploring characteristics of studies, participants, partners, interactions, and responses as moderators (see Methods for moderator coding). Of 23 response types meta-analysed, moderator analyses were not conducted for partner effects on prosocial behaviour, moral engagement, perceived self-agency, self-disclosure, strategic economic behaviour, and affective arousal due to non-significant heterogeneity in effect sizes ($Q$-test $p < 0.05$). Despite significant effect-size heterogeneity, the limited number of studies ($k < 10$)[69] precluded moderator analyses for perceived social presence, future interaction intention, perceived interaction naturalness and enjoyment, and subjective workload. Therefore, we conducted moderator analyses for 12 response types: perceived likeability, perceived competence, agency attribution, responsibility attribution, behavioural trust, subjective trust, social alignment, objective task performance, perceived partner relational qualities, affective valence, interaction satisfaction, and subjective task engagement.

For simplicity, this section only presents results for six response types (see Table 3) under the themes of "social perceptions of interaction partners" and "interaction experiences", with moderators both reaching frequentist significance and receiving at least substantial Bayesian support ($p < 0.05$; $BF_{10} > 3$). Full results are provided in Supplementary Table 4.

**Social perceptions of interaction partners.** For *perceived likeability*, partner appearance difference and interaction task moderated the partner effect. Agents were perceived as significantly less likeable than human partners (a medium partner effect) when appearance differed, but this was non-significant when appearances matched. Moreover, agents were significantly less likeable than human partners in service encounters and games (medium and high partner effects, respectively), but this was non-significant in instructional interactions and communication-focused interactions.

For *perceived competence*, appearance difference, agent form (physical vs. virtual), and interaction medium (face-to-face vs. computer-mediated) moderated the partner effect. Agents were perceived as significantly less competent than human partners (a medium-to-large partner effect) when appearance differed, but this was non-significant when appearances matched. The agent disadvantage manifested across conditions of agent form and interaction medium but was significantly greater for physical agents than virtual agents (medium-to-large vs. small) and greater in face-to-face than computer-mediated interactions (medium-to-large vs. small).

For *agency attribution*, interaction medium (face-to-face vs. computer-mediated) moderated the partner effect. Agents were attributed significantly less agency than human partners (a large partner effect) in face-to-face interactions; this partner effect was small-to-medium while non-significant in computer-mediated interactions.

For *responsibility attribution*, study setting (online vs. lab) and interaction realism (hypothetical vs. real-time)—which were perfectly co-linear—moderated the partner effect. In online studies with hypothetical interaction, agents were attributed significantly less responsibility than human partners (a medium-to-large partner effect), whereas lab studies with real-time interaction yielded a non-significant partner effect.

**Interaction experiences.** For *perceived partner relational qualities*, agent operationalisation (vignette-described vs. Wizard-of-Oz vs. autonomous), human partner type (vignette-described partner vs. research team member vs. algorithm-controlled pseudo-human), and interaction realism (hypothetical vs. real-time) moderated the partner effect. Perfect multicollinearity occurred among moderator conditions due to a subset of studies that asked participants to imagine interacting with partners (i.e., hypothetical interaction) through vignettes. This subset yielded a medium-to-large partner effect, indicating methodological contribution to agent disadvantage. Studies with real-time interaction yielded a non-significant partner effect, regardless of agent operationalisation or human partner type. Additionally, response dimension moderated this partner effect. Studies measuring perceived customer orientation showed that agents were perceived as significantly less customer-oriented than human partners (a medium-to-large partner effect), whereas studies of other relational qualities—perceived rapport, interactivity, and empathy—yielded non-significant partner effects.

For *interaction satisfaction*, interaction nature (oppositional vs. cooperative) moderated the partner effect. In oppositional interactions, satisfaction was significantly higher with agent vs. human partners (a medium partner effect), whereas in cooperative interactions, this was non-significant.

In brief, no participant characteristics demonstrated robust moderating effects (i.e., effects reaching frequentist significance with at least substantial Bayesian support). Nevertheless, significant moderators of partner effects on specific response types were identified across other categories: study characteristics (study setting), response characteristics (response dimension), partner characteristics (partner appearance difference, agent form, agent operationalisation, and human partner type), as well as interaction characteristics (interaction task, medium, realism, and nature).

**Publication bias, research quality, and sensitivity analysis**
For response types with at least ten studies[69], we evaluated publication bias via funnel plots and Egger Sandwich tests[70] and confirmed minimal risk of publication bias. Detailed results and interpretations are provided in Supplementary Fig. 1 and Supplementary Table 5. Research quality was assessed via a tailored checklist (Supplementary Table 6), yielding three quality metrics—study design rigour, data & reporting rigour, and broad research integrity—for each study (Supplementary Table 7). Meta-regressions showed no systematic impact of research quality on our results, though modest influences of specific quality metrics cannot be ruled out; details are provided in Supplementary Table 8. Finally, three sensitivity checks—incorporating approximated effect sizes, removing outliers and influential cases, and using alternative Bayesian priors—confirmed that our main meta-analytic results were robust across analytic decisions; details are provided in Supplementary Tables 9–11.

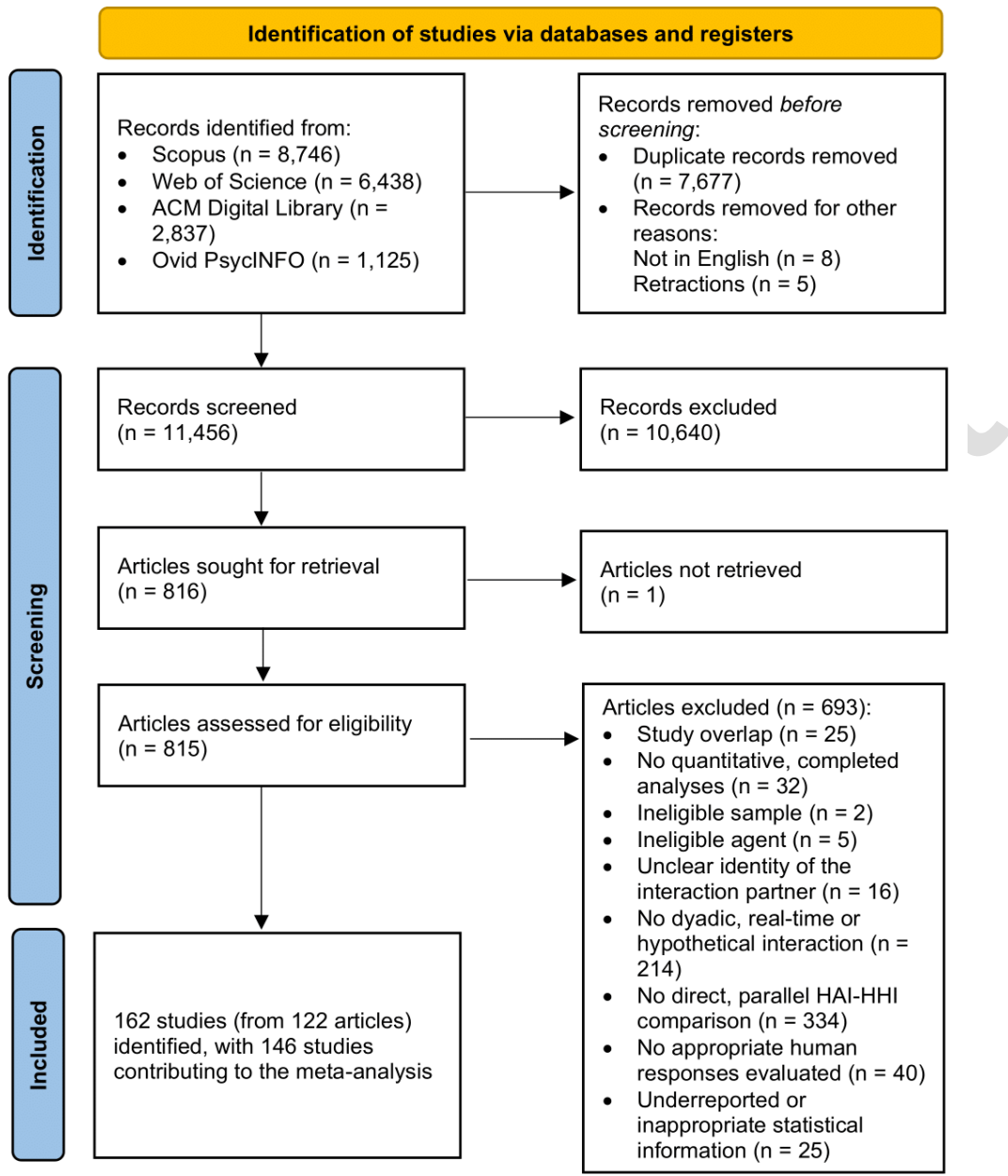

**Fig. 1. PRISMA flowchart**

*Note.* Two articles[71,72] reported the same underlying study while analysing different human responses, and were treated as one study in our meta-analysis. Articles[73,74] likewise reported the same study and were treated as one. There was one article[75] describing a single investigation but collecting and analysing data separately for Chinese and US samples; to keep sample independence, we treated these as two studies. Another article[76] reported Japanese and US samples, which we also treated as two studies. In addition, of the 25 articles excluded for underreported or inappropriate statistical information, 16 articles (22 studies) allowed deriving approximated effect sizes for some or all responses examined. These approximated effect sizes were included in the sensitivity analyses in the supplementary information.

**Table 1. Overview of the meta-analysed response types**

| Response themes | Responses | Conceptualised As |
|---|---|---|
| Prosociality and morality | Prosocial behaviour | Voluntary actions that are intended to help or benefit others[54]. |
| | Moral engagement | Psychological and behavioural commitment to moral standards in a given context, demonstrating direct or indirect responsiveness to the needs and interests of others[55,56]. |
| Social perceptions of interaction partners | Perceived social presence | Subjective experience of being present with a real social partner, with whom one can exchange thoughts and emotions[36,77]. |
| | Perceived likeability | Positive evaluation of the partner based on their affiliative capacity and social attractiveness[58,59]. |
| | Perceived competence | Evaluation of the partner's intelligence and domain-specific knowledge, or their effectiveness and efficiency in task performance[58,60]. |
| | Agency attribution | Explicit, reflective process of ascribing the partner the capacity to intentionally initiate events, and recognising them as the cause of their own decisions or actions[61]. |
| | Responsibility attribution | Process of assigning responsibility, with credit or blame, for an event or action to the partner[62]. It is closely related to agency attribution, which lays the foundation: the partner needs to be perceived as having agency to cause an outcome before being assigned responsibility for it[78,79]. |
| Trust in interaction partners | Behavioural trust | Observable actions of voluntarily accepting vulnerability based on positive expectations of the partner, such as risk-taking behaviour or reliance on the partner under uncertainty[80,81]. |
| | Subjective trust | Psychological state comprising the intention to accept vulnerability based on positive expectations of the partner, such as their ability, benevolence, or integrity[63,81]. |
| Social alignment with interaction partners | Social alignment | Voluntary alignment of internal states and behaviours with the partner, including self-other integration (merging identities and perspectives), advice taking (incorporating partner insights), synchrony (adapting verbal and nonverbal behaviours), and proximity regulation (regulating interpersonal distances)[64]. |
| Personal agency and task performance | Perceived self-agency | Subjective experience of controlling one's own body and external events, which leads one to feel responsible for what their decisions or actions cause[82]. |
| | Self-disclosure | Process of revealing personal information, such as one's own thoughts, feelings, and experiences, with the partner[83]. |
| | Strategic economic behaviour | Deliberate choices of actions in economic games where one, recognising the interdependence of actions, anticipates and reacts to the partner's actions by weighing how their choices would affect both personal and partner outcomes. It can be affected by multiple factors, including not only payoff maximisation, but also social preferences (e.g., reciprocity, trust, and fairness) or even intuitions[66,84]. |
| | Objective task performance | Measurable and quantifiable assessments regarding one's performance of specific tasks. |
| Interaction experiences | Perceived partner relational qualities | Subjective evaluations of the relational attributes the partner exhibits during interaction, including perceived rapport (closeness and mutual connection), perceived interactivity (active engagement and responsiveness), perceived empathy and supportiveness (understanding of and support for one's perspectives and feelings), and perceived customer orientation (commitment to meeting one's needs)[67]. |
| | Affective valence | Hedonic tone (i.e., unpleasantness–pleasantness) of the emotional experience when interacting with the partner[85]. |
| | Affective arousal | Activation level (i.e., calmness–excitement) of the emotional experience when interacting with the partner[85]. |
| | Interaction satisfaction | Overall evaluation of how well the interaction with the partner meets one's needs and expectations[68]. |
| | Future interaction intention | Degree to which one would like to interact with the partner again in the future[86]. |
| | Perceived interaction naturalness | Degree to which one perceives their interaction with the partner as natural. |
| | Perceived interaction enjoyment | Degree to which one feels they have enjoyed interaction with the partner. |
| | Subjective workload | Perception and emotional experience of the overall effort invested in performing specific tasks[87]. |
| | Subjective task engagement | Perception and emotional experience of involvement and investment in performing specific tasks[88]. |

Table 2. Meta-analytic results for different response types in human-agent vs. human-human interactions

| Response themes | Responses | Hedges' $g$ | 95% CI | $t$ | $t\_p$ | $Q$ | $Q\_p$ | $I^2$ (%) | $BF_{10}$ | $k$ | $m$ |
|---|---|---|---|---|---|---|---|---|---|---|---|
| Prosociality and morality | Prosocial behaviour | -0.648 | [-0.82, -0.48] | -8.66 | < 0.001 | 10.83 | 0.288 | 21.78 | 6003.065 | 9 | 10 |
|  | Moral engagement | -0.376 | [-0.48, -0.27] | -8.19 | < 0.001 | 4.04 | 0.991 | 0 | 1725.850 | 8 | 14 |
| Social perceptions of interaction partners | Perceived social presence | -0.284 | [-0.52, -0.05] | -2.81 | 0.023 | 48.72 | < 0.001 | 68.19 | 3.495 | 9 | 20 |
|  | Perceived likeability | -0.352 | [-0.54, -0.16] | -3.78 | < 0.001 | 443.07 | < 0.001 | 88.39 | 41.417 | 28 | 41 |
|  | Perceived competence | -0.457 | [-0.61, -0.30] | -6.21 | < 0.001 | 148.84 | < 0.001 | 77.56 | 10296.030 | 23 | 31 |
|  | Agency attribution | -0.705 | [-1.16, -0.25] | -3.43 | 0.006 | 229.52 | < 0.001 | 94.62 | 16.554 | 11 | 18 |
|  | Responsibility attribution | -0.511 | [-0.72, -0.30] | -5.44 | < 0.001 | 92.77 | < 0.001 | 81.56 | 512.117 | 12 | 18 |
| Trust in interaction partners | Behavioural trust | -0.020 | [-0.16, 0.12] | -0.30 | 0.769 | 47.98 | 0.002 | 60.83 | 0.077 | 18 | 24 |
|  | Subjective trust | -0.109 | [-0.23, 0.01] | -1.88 | 0.073 | 115.41 | < 0.001 | 76.87 | 0.358 | 25 | 35 |
| Social alignment | Social alignment | -0.004 | [-0.10, 0.09] | -0.08 | 0.937 | 52.90 | 0.027 | 33.11 | 0.052 | 22 | 36 |
| Personal agency and task performance | Perceived self-agency | 0.010 | [-0.14, 0.16] | 0.16 | 0.880 | 18.51 | 0.139 | 37.53 | 0.084 | 9 | 14 |
|  | Self-disclosure | 0.025 | [-0.16, 0.21] | 0.31 | 0.766 | 28.03 | 0.109 | 44.89 | 0.089 | 9 | 21 |
|  | Strategic economic behaviour | -0.086 | [-0.18, 0.01] | -1.93 | 0.078 | 32.82 | 0.084 | 32.26 | 0.246 | 13 | 24 |
|  | Objective task performance | 0.008 | [-0.10, 0.12] | 0.15 | 0.884 | 70.60 | < 0.001 | 46.39 | 0.062 | 23 | 38 |
| Interaction experiences | Perceived partner relational qualities | -0.157 | [-0.40, 0.09] | -1.36 | 0.194 | 159.82 | < 0.001 | 91.66 | 0.320 | 17 | 25 |
|  | Affective valence | -0.226 | [-0.57, 0.12] | -1.49 | 0.171 | 84.45 | < 0.001 | 86.33 | 0.410 | 10 | 14 |
|  | Affective arousal | -0.087 | [-0.34, 0.17] | -0.87 | 0.422 | 17.81 | 0.058 | 46.56 | 0.168 | 6 | 11 |
|  | Interaction satisfaction | 0.085 | [-0.12, 0.29] | 0.86 | 0.400 | 281.59 | < 0.001 | 94.73 | 0.154 | 19 | 27 |
|  | Future interaction intention | 0.105 | [-0.10, 0.31] | 1.32 | 0.245 | 14.10 | 0.028 | 57.34 | 0.212 | 6 | 7 |
|  | Perceived interaction naturalness | -0.368 | [-1.24, 0.50] | -1.18 | 0.304 | 73.40 | < 0.001 | 96.97 | 0.585 | 5 | 5 |
|  | Perceived interaction enjoyment | -0.182 | [-0.69, 0.32] | -0.88 | 0.411 | 83.97 | < 0.001 | 89.06 | 0.355 | 7 | 10 |
|  | Subjective workload | -0.157 | [-0.54, 0.23] | -1.13 | 0.323 | 32.49 | 0.002 | 69.04 | 0.320 | 5 | 14 |
|  | Subjective task engagement | -0.063 | [-0.31, 0.19] | -0.57 | 0.585 | 20.85 | 0.022 | 58.66 | 0.168 | 10 | 11 |

*Note.* For simplicity, this table presents results primarily from frequentist meta-analysis. $BF_{10}$ is presented to provide complementary Bayesian evidence for $H_1$ over $H_0$. All frequentist estimates are consistent with Bayesian estimates. Full Bayesian results are presented in Supplementary Table 12. Specifically, Hedges' $g$, 95% CI, $t$-value and associated $p$-value were estimated via random-effects meta-analysis. $Q$ is Cochrane's $Q$-statistic for testing heterogeneity. $I^2$ is the proportion of total variance attributable to true heterogeneity rather than sampling error. $k$ is the number of studies including in meta-analysis. $m$ is the number of effect sizes included.

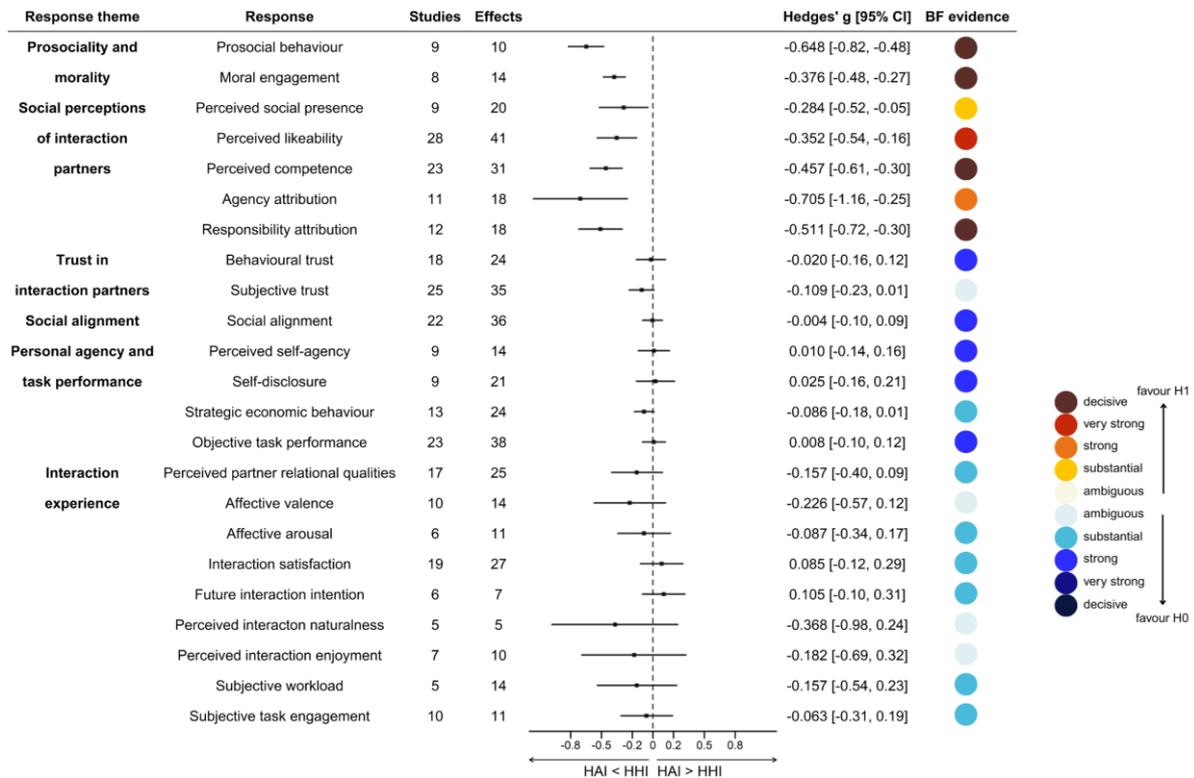

**Fig. 2. Forest plot visualising pooled effect sizes for different response types**

*Note.* Only response types with five or more studies were included in the meta-analysis and are shown in this plot. Horizontal bars represent 95% CIs. Response types with bars that do not cross the dashed vertical "line of no effect" indicate existing significant differences between human-agent and human-human interactions. Marks on the x-axis denote conventional thresholds for small ($g = 0.2$), medium ($g = 0.5$), and large ($g = 0.8$) effects. The pooled effect sizes shown in this plot were derived from frequentist meta-analyses, with the accompanying strength of Bayesian evidence indicated by coloured circles.

Table 3. Meta-regression and subgroup analysis results for different response types

| Responses | Meta-regressions | Moderator conditions | Hedges' g | 95% CI | t | t_p | k | m |
|---|---|---|---|---|---|---|---|---|
| Perceived likeability | Appearance difference: $F = 7.48$, $p = 0.011$; $BF_{10} = 4.559$ | Appearance differed | -0.539 | [-0.81, -0.27] | -4.23 | < 0.001 | 17 | 27 |
| | | Appearance matched | -0.062 | [-0.27, 0.15] | -0.66 | 0.524 | 11 | 14 |
| | Interaction task: $F = 5.16$, $p = 0.007$; $BF_{10} = 4.839$ | Service encounter | -0.587 | [-0.87, -0.30] | -4.74 | 0.001 | 9 | 11 |
| | | Game | -0.825 | [-1.19, -0.46] | -7.16 | 0.006 | 4 | 5 |
| | | Instructional interaction | 0.101 | [-0.62, 0.82] | 0.44 | 0.688 | 4 | 5 |
| | | Communication-focused | -0.130 | [-0.50, 0.24] | -0.81 | 0.440 | 9 | 17 |
| Perceived competence | Agent form: $F = 13.42$, $p = 0.001$; $BF_{10} = 16.598$ | Physical | -0.669 | [-0.82, -0.52] | -9.69 | < 0.001 | 13 | 16 |
| | | Virtual | -0.233 | [-0.44, -0.02] | -2.49 | 0.035 | 10 | 14 |
| | Appearance difference: $F = 12.49$, $p = 0.002$; $BF_{10} = 19.646$ | Appearance differed | -0.614 | [-0.76, -0.47] | -8.91 | < 0.001 | 15 | 22 |
| | | Appearance matched | -0.170 | [-0.43, 0.09] | -1.54 | 0.167 | 8 | 9 |
| | Interaction medium: $F = 12.08$, $p = 0.002$; $BF_{10} = 14.841$ | Face-to-face | -0.712 | [-0.86, -0.57] | -11.48 | < 0.001 | 9 | 12 |
| | | Computer-mediated | -0.275 | [-0.48, -0.07] | -2.87 | 0.014 | 13 | 18 |
| Agency attribution | Interaction medium: $F = 7.09$, $p = 0.026$; $BF_{10} = 4.331$ | Face-to-face | -1.254 | [-2.33, -0.18] | -3.72 | 0.034 | 4 | 6 |
| | | Computer-mediated | -0.367 | [-0.79, 0.05] | -2.14 | 0.077 | 7 | 12 |
| Responsibility attribution | Study setting: $F = 11.34$, $p = 0.007$; $BF_{10} = 4.271$ | Online | -0.630 | [-0.85, -0.41] | -6.51 | < 0.001 | 9 | 11 |
| | | Lab | -0.188 | [-0.59, 0.21] | -2.02 | 0.181 | 3 | 7 |
| | Interaction realism: $F = 11.34$, $p = 0.007$; $BF_{10} = 4.271$ | Hypothetical | -0.630 | [-0.85, -0.41] | -6.51 | < 0.001 | 9 | 11 |
| | | Real-time | -0.188 | [-0.59, 0.21] | -2.02 | 0.181 | 3 | 7 |
| Perceived partner relational qualities | Agent operationalisation: $F = 11.37$, $p = 0.001$; $BF_{10} = 71.227$ | Vignette-described | -0.709 | [-1.20, -0.22] | -4.01 | 0.016 | 5 | 6 |
| | | Wizard-of-Oz | 0.123 | [-0.14, 0.39] | 1.10 | 0.308 | 8 | 15 |
| | | Autonomous | 0.024 | [-0.17, 0.22] | 0.39 | 0.726 | 4 | 4 |
| | Human partner type: $F = 10.75$, $p = 0.002$; $BF_{10} = 45.469$ | Vignette-described partner | -0.709 | [-1.20, -0.22] | -4.01 | 0.016 | 5 | 6 |
| | | Pseudo-human | -0.004 | [-0.32, 0.31] | -0.06 | 0.958 | 3 | 3 |
| | | Research team member | 0.118 | [-0.13, 0.36] | 1.15 | 0.289 | 8 | 15 |
| | Interaction realism: $F = 23.50$, $p < 0.001$; $BF_{10} = 298.645$ | Hypothetical | -0.709 | [-1.20, -0.22] | -4.01 | 0.016 | 5 | 6 |
| | | Real-time | 0.081 | [-0.06, 0.23] | 1.23 | 0.243 | 12 | 19 |
| | Response dimension: $F = 6.78$, $p = 0.006$; $BF_{10} = 3.334$ | Perceived customer orientation | -0.873 | [-1.31, -0.44] | -8.59 | 0.013 | 3 | 3 |
| | | Perceived rapport | 0.171 | [-0.88, 1.23] | 0.70 | 0.557 | 3 | 10 |
| | | Perceived interactivity | -0.003 | [-0.12, 0.11] | -0.05 | 0.960 | 6 | 6 |
| | | Perceived empathy | -0.218 | [-0.65, 0.21] | -1.00 | 0.374 | 5 | 5 |
| Interaction satisfaction | Interaction nature: $F = 40.31$, $p < 0.001$; $BF_{10} = 15137.576$ | Oppositional | 0.493 | [0.23, 0.76] | 4.77 | 0.005 | 6 | 6 |
| | | Cooperative | -0.081 | [-0.29, 0.13] | -0.83 | 0.421 | 15 | 20 |

*Note.* Moderator analyses were conducted for response types with significant effect-size heterogeneity and at least ten available studies. Potential moderators were tested individually. For categorical moderators with multiple conditions, only conditions represented by at least three studies were included. For simplicity, this table presents results only for moderators that both reached frequentist significance and received at least substantial Bayesian support ($p < 0.05$; $BF_{10} > 3$). Full results for all tested moderators are available in Supplementary Table 4. For meta-regression results, $F$-value is from random-effects meta-regression. $BF_{10}$ represents the Bayesian evidence for $H_1$ (i.e., presence of a moderating effect) over $H_0$ (i.e., absence of a moderating effect). For follow-up subgroup analysis results, Hedges' $g$, 95% CI, $t$-value and associated $p$-value were estimated via random-effects meta-analysis. $k$ is the number of studies including in the analysis. $m$ is the number of effect sizes included.

**Discussion**
This review synthesised empirical evidence on similarities and differences in individuals' psychological and behavioural responses when interacting with performance-matched agent vs. human partners. We conducted separate random-effects meta-analyses for 23 types of human responses across six themes, followed by univariate meta-regressions for 12 response types with significant effect-size heterogeneity and at least ten available studies.

Our meta-analyses found that individuals were less prosocial and morally engaged when interacting with agents than with human partners. These findings extend and reinforce a prior review showing increased selfishness and rationality when playing behavioural economic games with computer vs. human partners[89]. Additionally, compared to human partners, on average, agents were perceived as possessing fewer social attributes: they were rated as less likeable and competent, and attributed less agency and responsibility. Agents were also perceived to have lower social presence. However, this partner effect was small and sensitive to how missing non-significant effects were handled, suggesting that more studies are needed before reaching a robust conclusion about perceived social presence.

Alongside response differences, we identified key response similarities between HAI and HHI. Specifically, individuals demonstrated social alignment and behavioural trust with agent and human partners comparably. Subjective trust likewise showed no significant difference in frequentist meta-analysis, whereas weak Bayesian support for the null effect leaves open the possibility of a modest partner effect. In addition, individuals exhibited comparable self-agency and self-disclosure in both interactions. Apart from these self-oriented processes, objective task performance and strategic economic behaviour were also comparable between HAI and HHI. Finally, most interaction experiences were on average similar between partner types. Bayesian support for no partner effects was ambiguous for affective valence and perceived interaction naturalness and enjoyment, so modest partner effects for these three responses cannot be ruled out.

Furthermore, many subjective responses (i.e., social perceptions of partners, subjective trust, and interaction experiences) exhibited high effect-size heterogeneity, indicating that the average effects from meta-analyses should not be interpreted as universal, with true partner effects being context-dependent. Across examined social attributes, agents were on average perceived less positively than human partners, but several moderators shaped these partner effects. Specifically, the partner effect on perceived likeability disappeared when partner appearance matched (vs. differed) or in instructional/communication-focused interaction (vs. game/service encounter) tasks. The partner effect on perceived competence disappeared when partner appearance matched (vs. differed) and was reduced when agents were virtual (vs. physical) or in computer-mediated (vs. face-to-face) interactions. The partner effects on agency and responsibility attribution disappeared in computer-mediated (vs. face-to-face) interactions and in lab-setting, real-time (vs. online-setting, hypothetical) interactions, respectively. Given the limited studies for many interaction experiences, we identified moderators only for interaction satisfaction and perceived partner relational qualities. Interaction satisfaction was higher with agents than with human partners in oppositional (vs. cooperative) interactions. Agents were perceived as having lower relational qualities than human partners in hypothetical, vignette-described (vs. real-time) interactions, or when focusing on perceived customer orientation (vs. interactivity/rapport/empathy). For many other tested moderators, evidence regarding their influence was inconclusive—for example, some reached frequentist significance but with weak Bayesian support (e.g., measurement timing for subjective trust)—suggesting more research is needed to clarify variations in these subjective responses. Overall, we believe the partner effects on these responses should be interpreted by considering both the average effects and the strong effect-size heterogeneity.

These findings should be interpreted considering the review limitations. First, despite all studies being peer-reviewed, research quality varied greatly, with three quality metrics averaging only moderate levels. While we found no systematic impact of research quality on results, missing or inconsistent statistics and methodological details in some studies could add noise to certain meta-analytic estimates. Second, our response classification was derived inductively from the variables and measures reported in the included studies and thus cannot capture the full spectrum of human responses possible in HAI/HHI. Third, we restricted moderator analyses to univariate models. Multivariate models, even Bayesian regularised meta-regression[90], would be uncertain due to lack of sufficient studies to accommodate multiple moderators simultaneously. Additionally, many subjective responses exhibited high effect-size heterogeneity, which our meta-regressions only partly explained. Measurements used across studies varied in format, reliability, and validity, but they were too diverse to be fully captured in our moderator coding, potentially obscuring nuanced moderating effects.

Fourth, our findings are limited by the scope of this review. We focused on responses in dyadic interactions, so generalising our partner-effect findings to multi-party scenarios requires caution. We analysed individual-level responses directly related to the interaction, whereas dyad-level responses and those reflecting downstream outcomes were beyond scope. Moreover, most included studies investigated one-time interactions in controlled lab or online settings; responses in sustained, naturalistic interactions remain unclear. Future work should revisit partner effects using more sophisticated agents capable of long-term interaction. This was partly due to earlier technological constraints, with agents in most included studies relying on traditional machine learning, rule-based algorithms, or Wizard-of-Oz setups. Although our inclusion of "intelligent agent" was technology-agnostic, covering systems from traditional technologies to modern GenAI, the GenAI-based studies we initially identified did not meet our study eligibility criteria. Nonetheless, we expect that GenAI's growing prevalence will spur new comparisons of interaction with GenAI agents vs. humans, enabling re-examination and extension of the current findings. In addition, our review covered only general adult populations, and was dominated by WEIRD samples[91], with limited non-WEIRD representation mainly from East Asia. Future reviews should target younger populations and include broader non-WEIRD populations by synthesising non-English studies.

Despite its limitations, this review demonstrates several methodological strengths. It is the first meta-analysis to compare a wide range of individuals' psychological and behavioural responses when interacting with performance-matched agent vs. human partners. We rigorously classified the diverse responses using an empirical-to-conceptual approach[53]. Alongside frequentist analyses, we conducted Bayesian meta-analyses to confirm result consistency. The Bayesian approach also provided complementary insights: Bayes factors quantified the strength of evidence for effect presence/absence, helping distinguish null effects from underpowered results and flag ambiguous cases where pooled effects warrant caution. Sensitivity checks—excluding outliers and influential cases, incorporating approximated effect sizes, and testing alternative Bayesian priors—confirmed the robustness of our main results across analytic decisions. We further tested many potential moderators to explore sources of significant effect-size heterogeneity. Finally, recognising that existing research appraisal tools were less suited to our review, we developed a tailored checklist to assess research quality while capturing its multidimensional nature.

This review also has significant theoretical implications. We synthesised empirical evidence across HCI, social robotics, psychology, communication, and business studies, establishing a holistic understanding of similarities and differences in human responses between HAI and HHI. By spanning diverse agent morphologies (e.g., robots, virtual humans, conversational agents), we found that, except for social perceptions and interaction experiences, partner effects on most responses were consistent regardless of agent form (physical vs. virtual) or embodiment (disembodied vs. embodied). This suggests that the cognitive mechanisms underlying core collaborative processes like trust and social alignment may be morphology-agnostic, though this warrants further investigation. Moreover, our results confirmed that social responses, including social perceptions, prosocial behaviour, moral engagement, trust, and social alignment, should be considered as distinct constructs rather than a monolithic entity[42]. We suggest expanding existing theoretical frameworks (e.g., Media Equation[29], Threshold Model of Social Influence[31]) by specifying which types of social responses converge or diverge between HAI and HHI and identifying moderators for these patterns. Overall, our review benchmarks human responses in HAI vs. HHI during the pre-GenAI era and provides a reference for future human-GenAI research, enabling researchers to detect potential shifts in human psychology and behaviour as agent technologies evolve.

The findings from this review have implications for the development of interactive intelligent agents. First, we found that behavioural trust, social alignment, personal agency, and objective task performance were comparable in HAI and HHI. That is, when assuming human-equivalent roles with matched performance, agents elicit behavioural trust, facilitate effective interaction, and preserve user agency and performance. This indicates that well-performing agents are afforded instrumental value on par with humans, making them promising collaborators in goal-directed tasks. To move beyond instrumental parity towards successful collaboration, our review also reveals key design considerations. Specifically, we found that agents were attributed less responsibility than humans, whereas collaboration typically assumes shared responsibility[92], raising concerns that individuals paired with agents may be overburdened with accountability. This necessitates designing robust accountability structures and clear human-agent communication. Another design consideration arises from our review's focus on agents with human-level performance, which may promote reciprocity and safeguard user agency. However, there are scenarios where leveraging agents' superhuman capabilities and granting them primary control may be advantageous. Recent AI advances have enabled these agents to surpass humans in both knowledge-based tasks[2,4,5] and situational judgment tasks[93]. Thus, agent design should strategically calibrate agent capabilities to task demands while ensuring transparency and explainability.

Second, we found reduced prosocial behaviour and moral engagement with agent vs. human partners. Agents were also perceived as less competent, likeable, socially present, and attributed less agency and responsibility. These deficits indicate that agents are not afforded intrinsic value at a level compared to humans. Prior research showed that participants compensated an ostracised agent during a Cyberball game[94], mirroring behaviour observed in human-human ostracism; yet, the compensation effect for agents was smaller than the large effect reported for human targets[95,96]. Jointly, this suggests that agents receive some intrinsic value but at meaningfully lower levels. This gap warrants caution in morally sensitive domains like healthcare, education, and social services. Developers should ensure system benefits outweigh risks of diminished user prosociality and morality and retain human oversight for critical decisions. Design efforts should be put in promoting prosociality/morality in HAI, potentially leveraging psychological levers like psychological flexibility[97] and gratitude[98]. Regulatory infrastructure, including robust accountability frameworks, mandatory incident logging[99], and harm mitigation protocols, should be embedded throughout development cycles.

Finally, whereas agents were generally perceived as possessing fewer social attributes, high heterogeneity in these partner effects highlights important design opportunities. Our review uncovered several agent and interaction characteristics that moderated the partner effects on different social perceptions, which indicates that these agent disadvantages are malleable through strategic design and deployment. Developers should consider incorporating cost-effective social cues, optimising interaction contexts, and calibrating designs to task-specific social requirements. Likewise, high effect-size heterogeneity for subjective trust and interaction experiences warrants further moderator investigation to enable targeted design efforts.

**Methods**
**Literature search and eligibility criteria**
This systematic review and meta-analysis followed the Preferred Reporting Items for Systematic Reviews and Meta-Analyses (PRISMA) 2020 statement. We conducted systematic searches in February 2024 across four literature databases: Scopus (which covers IEEE Xplore[100]), Web of Science, ACM Digital Library, and PsycInfo. We searched titles, abstracts, and keywords using search strings comprising two components. The first component identified various types of interactive intelligent agents, while the second combined general and specific terms to capture a wide range of HHI topics, including common interpersonal and social phenomena that have been studied in HAI. The searches focused on empirical research articles published from 2000 and employing quantitative or mixed methods (extracting only quantitative data) and were restricted to peer-reviewed journal or conference papers written in English. The full search strings can be found in Supplementary Table 13. Grey literature was not searched due to challenges in locating studies and the lack of a reliable method for assessing quality.

We defined the eligibility criteria via the PECO (Population, Exposure, Comparison, and Outcome) framework. Eligible studies were on healthy adult participants (P) that investigated both dyadic HAI (E) and HHI (C) and made direct, parallel comparisons of human responses (O) in these two conditions. Interactive intelligent agents in HAI could be either physical or virtual, embodied or disembodied. Participants actively engaged in research tasks that involved (perceived) real-time or hypothetical interactions with partners, whom participants believed to be either agent or human. Additionally, participants received similar treatment in HAI and HHI; that is, the interaction task, dynamics, and partner role and performance were consistent across both conditions, except for the partner's voice and appearance. When voice and appearance were matched between agent and human partners, the only difference lay in the perceived identity of the interaction partner, isolating the identity effects on human responses. However, as voice and appearance were often used to enhance the believability of the manipulated partner identity, variations in these two aspects were permitted. Lastly, human responses were examined in terms of participants' individual-level psychological and behavioural responses during or after the interactions.

The detailed inclusion and exclusion criteria are provided in Supplementary Table 1. The study selection involved two stages: title-abstract screening and full-text screening. At each stage, the first author screened all the candidate articles, and two co-authors each screened a random 15% subset. Any discrepancies were resolved through discussion.

**Data extraction and coding**

The data extraction and coding protocol (including study details and essential statistics for effect size calculation; Supplementary Table 14) was established a priori, except that human response classification and the specific levels of two moderators (i.e., interaction task and response dimension) were determined post hoc. The first author completed data extraction and coding for all eligible studies, and two co-authors each checked a random 15% subset. Any discrepancies were resolved through discussion. Due to the diversity of human responses examined across studies, we performed an a posteriori classification and meta-analysed each response type separately. We initially applied a conceptual-to-empirical approach[53], but the extracted responses did not align with the responses collected in Krpan's taxonomy[23]. We thus shifted to an empirical-to-conceptual approach[53], inductively classifying responses into distinct types and then iteratively grouping conceptually aligned response types into overarching themes.

In addition, we identified five categories of potential moderators to explore sources of potential heterogeneity in effect sizes: study characteristics, participant characteristics, partner characteristics, interaction characteristics, and response characteristics.

*Study characteristics.* Study setting referred to the environment in which the study was conducted, coded as laboratory, online, or field. Study design captured the assignment of participants to HAI and HHI conditions, coded as between-subjects or within-subjects. As only a few studies reported data collection time, publication year was extracted as a proxy continuous moderator to explore potential temporal trends in effect sizes. Publication type was coded as journal or conference.

*Participant characteristics.* Participants in the eligible studies were drawn from diverse countries. Sample location was coded by continent: Asia, Europe, North America, or Oceania (no studies from Africa or South America were identified). The sample was also coded as WEIRD (Western, Educated, Industrialised, Rich, Democratic) or non-WEIRD countries[91]. The average age of participants and the percentage of female participants were extracted as continuous moderators.

*Partner characteristics.* This category included characteristics of both partners that might affect participants' responses. Human partner type was coded as research team member (experimenter, research assistant or hired actor/specialist following specific interaction scripts), participant (untrained individual), pseudo-human (ostensibly human partner controlled by algorithms), or vignette-described partner (human partner and their behaviour described through vignettes). Agent partner operationalisation during the interaction was coded as autonomous (behaving independently via algorithms), Wizard-of-Oz (partially or fully controlled by hidden human operators), or vignette-described. Agent form was coded as virtual (existing within digital interfaces) or physical (tangible in the physical world). Agent embodiment was coded as disembodied (no visible presence, e.g., bots) or embodied (visibly represented, e.g., robots and virtual humans). For studies involving robots, robot appearance human-likeness was coded into four levels based on the ABOT Database/Predictor[101]: non-humanoid (scored 0–10), semi-humanoid (10–40), humanoid (40–70), or android (70–100). Furthermore, appearance difference and voice difference were binary-coded to indicate whether these two features differed between agent and human partners.

*Interaction characteristics.* Interaction realism was coded as real-time (actual or perceived real-time engagements with partners) or hypothetical (vignette-based imaginative interactions). Interaction flow was coded as bidirectional (mutual exchanges with partners) or unidirectional (one-sided input). Interaction medium referred to the channel through which the interaction occurred, coded as face-to-face, computer-mediated, or VR-mediated. Interaction structure captured the extent to which the interaction followed a predefined or scripted format, coded as non-structured, semi-structured, or structured. Interaction nature captured the goal alignment and relational dynamics shaping exchanges between participants and partners, coded as neutral, cooperative, oppositional, or mixed. Interactant power symmetry was coded as symmetrical (equal power or status, e.g., fellow game players) or asymmetrical (power imbalance, e.g., customer and service employee dyad). Unlike other moderators with predefined levels, interaction tasks were classified inductively based on emergent themes observed across studies: service encounter (e.g., customer service), game interaction (e.g., economic games), instructional interaction (e.g., tutoring), communication-focused interaction (e.g., dialogues and question-answer exchanges without explicit service provision, game mechanics, or instructional objectives), motor task (e.g., physical coordination), or other (unclassified).

*Response characteristics.* Human responses were broadly coded as psychological (subjective perceptions and experiences) or behavioural (observable behaviours and performance metrics). Measurement timing captured when responses were recorded, coded as during (e.g., in-task behaviour) or after the interaction (e.g., post-interaction scale). Finally, where applicable, some response types were further classified into different dimensions (i.e., subconstructs), and this variation was tested as a post hoc moderator.

**Research quality assessment**

We assessed the research quality of each eligible study, as its quality can bias effect sizes and higher-quality studies generally yield findings that more closely converge on the truth[102]. The first author completed quality assessment for all studies, and two co-authors each checked a random 15% subset. Any discrepancies were resolved through discussion.

Existing quality assessment tools, however, posed challenges. Widely used ones (e.g., RoB-2[103] and ROBIS[104]) were developed for health-related intervention research and tailored to specific methodological designs (like randomised controlled trials). These tools thus place great emphasis on assessing methodological features such as participant allocation and blinding that are less standard in HCI studies. Other popular tools are easy to apply but lack critical assessment criteria such as sample size and study design (e.g., MMAT[105] and JBI checklists[106]), or are generalised for mixed- and multi-method studies, with criteria too broad to allow robust assessment of quantitative studies in a meta-analysis (e.g., QuADS[107] and QualSyst[108]). Therefore, although developing a research quality assessment tool was not the primary goal of this review, we decided to devise a tailored checklist by adapting items from existing popular tools[103–108] and drawing on prior work that customised quality assessment tools for their reviews[109,110]. This 23-item quality checklist is available in Supplementary Table 6.

To further test the impact of research quality on the pooled effect sizes, we accounted for its multidimensional nature rather than collapsing items into a single score[69,111]. Specifically, we computed three composite quality metrics. First, *study design rigour* reflects how thoughtfully each study was planned and designed, assessed through quality items on objectives and preregistrations, participants, and study design. Second, *data & reporting* *rigour* reflects how rigorously study data were handled, assessed through items on data collection, analysis, and results. Third, *broad research integrity* captures broader practices that support reproducibility and trustworthiness, assessed through items on discussion, ethics, and open science. The first two metrics directly addressed a study's risk of bias, while the third related more to the overall research quality.

**Data analysis**

All analyses were run in R v4.2.2. For each eligible study, we calculated the standardised mean difference between HAI and HHI for specific human responses, serving as the effect size estimates in subsequent analyses. Hedges' *g* was chosen over Cohen's *d* to account for the potential bias in estimating effects with small sample sizes. Hedges' *g* of 0.1 was interpreted as a negligible effect size, 0.2 as small, 0.5 as medium, and 0.8 as large[112]. When required statistics for calculating unadjusted effect sizes were missing or contained substantial errors, we contacted corresponding authors to request clarification and essential statistical details or raw data, as per their preference. If authors did not respond or provide the requested information, we excluded those missing effects from the main analyses (and, if a study lacked all effects, we excluded that study in full). Nevertheless, when it was possible to approximate missing effects, either by estimating from reported thresholds (e.g., assuming $p = 0.005$ for $p < 0.005$), deriving statistically adjusted estimates, or imputing non-significant effects (i.e., *max$^+$*, *max$^-$*, and *zero*-coded[37,41]), we retained those approximated effect sizes in sensitivity analysis. Supplementary Table 15 presents all formulae for effect size calculation.

We conducted separate meta-analyses for different types of human responses, proceeding with a specific response type only when five or more studies were included[50]. Given that the included studies were expected to differ in samples, design, and measures, we employed random-effects meta-analysis. This approach accounts for both sampling error and between-study heterogeneity in effect sizes, yielding a pooled effect size that represents the mean of a distribution of true effects rather than a single fixed effect[113]. Thus, the random-effects framework ensures that the meta-analytic results can be generalised to a broader "population" of potential studies beyond those included in the analysis. Moreover, to accommodate dependencies between multiple effect sizes per response outcome within studies, we employed a three-level random-effects model. Three sources of variance were identified: random sampling error (level 1), variance among effect sizes within studies (level 2), and variance among effect sizes between studies (level 3)[114]. We applied both frequentist and Bayesian approaches in the meta-analysis.

We conducted frequentist meta-analysis via the *metafor* R package[115], employing a restricted maximum likelihood estimator to model heterogeneity and Knapp-Hartung adjustment to calibrate the standard error of the pooled effect size. We assessed effect-size heterogeneity using two indicators: (1) *Q*-statistic, which tests for the presence of heterogeneity via the significance of its p-value, and the (2) $I^2$-statistic, which quantifies the magnitude of the heterogeneity. $I^2$ represents the proportion of total variance among effect sizes attributable to true heterogeneity rather than sampling error, with values of 25%, 50%, and 75% interpreted as low, moderate, and high heterogeneity[116]. The presence of heterogeneity indicates the need for further moderator analysis (i.e., meta-regression) to explore potential sources of this heterogeneity. Each potential moderator was evaluated individually only when data from at least ten studies were available[69]. For categorical moderators with multiple conditions, only conditions represented by at least three studies were included[117]. Meta-regression was also employed to test whether research quality systematically impacted the pooled effect size, with three quality metrics evaluated individually. Additionally, we checked publication bias for response types with at least ten studies[69] by visually inspecting funnel plots for asymmetry and performing Egger Sandwich tests via the *clubSandwich* R package[118], with regression slope significance serving as a statistical indicator of asymmetry. We further conducted sensitivity analysis for outliers (effect sizes with absolute studentised deleted residuals exceeding 1.96[119]) and influential cases (identified using Cook's distance, DFBETAS values, and hat values[115]), re-running all meta-analyses after removing these cases to assess result robustness.

To complement the traditional frequentist meta-analytic approach, we also conducted Bayesian meta-analysis via the *brms* and *bridgesampling* R packages[120,121]. Bayesian analysis allows incorporation of prior information to better estimate different sources of variance and enables direct probability statements about parameters via credible intervals[122]. We pre-determined prior distributions for effect size estimates and heterogeneity parameters. For effect sizes, we used a Cauchy(0, $1/\sqrt{2}$) distribution, which has become a default prior in the field of psychology[123], to reflect uncertainty about both the direction and magnitude of effect sizes when comparing responses in HAI vs. HHI. For between-study variance, we used an inverse-Gamma(1, 0.15) distribution, based on between-study heterogeneity estimates from meta-analyses reported in *Psychological Bulletin* from 1990-2013[123,124]. For within-study variance, we used an inverse-Gamma(1, 0.1) distribution, reflecting a priori expectation that variance among effect sizes within studies is smaller than variance between studies. We used a Cauchy(0, $1/\sqrt{2}$) distribution for moderator effects. We also assessed the sensitivity of Bayesian meta-analytic results to prior distributions by using the following alternatives. For effect sizes, we considered a Student-*t*(3, 0, 1) prior, a weakly informative distribution that places less density at zero and more on moderate-to-large effects and has lighter tails than Cauchy(0, $1/\sqrt{2}$). We considered half-Cauchy(0, 0.3), a commonly used weakly informative prior[125], for between-study variance and half-Cauchy(0, 0.2) for within-study variance.

To ensure model validity, we performed Markov Chain Monte Carlo diagnostics: potential scale reduction statistics ($\hat{R} \leq 1.01$) for convergence[126], and effective sample sizes (ESS $\geq$ 1,000) for all parameters for sampling efficiency[120]. In addition, the Bayesian approach provides formal measures (i.e., Bayes factors; $BF_{10}$) of the strength of evidence for the study hypothesis ($H_1$) relative to the null hypothesis ($H_0$), thereby indicating when pooled effects should be interpreted with caution[112]. Specifically, $BF_{10} < 1$ indicates the data are more supportive of $H_0$ than $H_1$, with values of 1/3–1 indicating ambiguous evidence, 1/10–1/3 substantial evidence, 1/30–1/10 strong evidence, 1/100–1/30 very strong evidence, and < 1/100 decisive evidence in support of $H_0$; $BF_{10} = 1$ indicates perfect ambiguity; $BF_{10} > 1$ indicates the data are more supportive of $H_1$, with values of 1–3, 3–10, 10–30, 30–100, and > 100 indicating ambiguous, substantial, strong, very strong, and decisive evidence in support of $H_1$, respectively[127].

**Data availability**
This review was not preregistered. Effect sizes, moderator values, and research quality ratings for individual studies, along with R code and materials for reproducibility, are available via OSF.io (https://osf.io/4x26r/?view_only=352d4fe5a88c4b41b80fe5675c085bf4).

**Acknowledgements**
We sincerely thank Dr Miguel Vadillo for his guidance on meta-analytical methodology and research quality checklist development, and Dr Brady Roberts and Dr David B. Wilson for their guidance on effect size calculations. We are also grateful to Imperial College London librarians Katharine Thompson and Nicole Urquhart for their support with EndNote and the literature search in the early stages of the review. Finally, we extend our sincere thanks to all the authors who responded to our inquiries and shared their data and statistical information with us despite busy schedules.